\documentclass{article}
\usepackage{spconf,amsmath,graphicx}
\usepackage{cite}
\usepackage{amsfonts}
\usepackage{multirow}

\usepackage{enumitem}
\setlist{nosep, leftmargin=14pt}
\usepackage[colorlinks, citecolor=green]{hyperref}

\usepackage{mwe} 

\title{Modeling Causal Interactions Across Brain Functional Subnetworks for Population-specific Disease Analysis}
\name{\shortstack{
Alissen Moreno$^{1}$, Yingying Zhang$^{1}$, Qi Huang$^{2}$, Fabian Vazquez$^{1}$, Jose A. N\'u\~nez$^{1}$, Erik Enriquez$^{1}$, \\
Dongchul Kim$^{1}$, Kaixiong Zhou$^{3}$, Hongchang Gao$^{4}$, Pengfei Gu$^{1}$, Liang Zhan$^{5}$, Haoteng Tang$^{1,*}$ \thanks{$*$ Corresponding Author}
\thanks{Submitted and Reviewed by IEEE-ISBI 2026}
}}
\address{1. Department of Computer Science, University of Texas Rio Grande Valley, Edinburg, TX\\
2. Mallinckrodt Institute of Radiology, Washington University in St. Louis, St. Louis, MO\\
3. Department of Electrical and Computer Engineering, North Carolina State University, Raleigh, NC\\
4. Department of Computer and Information Sciences, Temple University, Philadelphia, PA\\
5. Department of Electrical and Computer Engineering, University of Pittsburgh, Pittsburgh, PA}

\begin{document}

\maketitle
\begin{abstract}
Current neuroimaging studies on neurodegenerative diseases and psychological risk factors have been developed predominantly in non-Hispanic White cohorts, with other populations markedly underrepresented.  
In this work, we construct directed hyper-connectomes among large-scale functional brain systems based on causal influences between brain regions, and examine their links to Alzheimer’s progression and worry levels across racial groups.
By using Health and Aging Brain Study–Health Disparities (HABS-HD) dataset, our experimental results suggest that neglecting racial variation in brain network architecture may reduce predictive performance in both cognitive and affective phenotypes.
Important shared and population-specific hyper-connectome patterns related to both AD progression and worry levels were identified.
We further observed distinct closed-loop directed circuits across groups, suggesting that different populations may rely on distinct feedback-based network regulation strategies when supporting cognition or managing emotional states.
Together, these results indicate a common backbone of network vulnerability with population-dependent variations in regulatory coordination, underscoring the importance of population-aware neuroimaging models.
\end{abstract}
\begin{keywords}
Racial heterogeneity; Early Alzheimer’s disease; Functional brain subsystems; Directed network interactions; Worry-related neurodynamics
\end{keywords}

\section{Introduction}
\label{sec:intro}
Recent advances in data-driven artificial intelligence (AI) applied to neuroimaging have greatly enhanced early diagnosis of neurodegenerative diseases and improved mental health prediction.
However, racial diversity remains severely underrepresented in this field, despite mounting evidence that both neurodegenerative progression and psychiatric conditions vary across racial groups \cite{babulal2019perspectives, sterling2022demographic, tang2025brain}. 
This gap stems largely from uneven recruitment practices, with racially minoritized populations (e.g., Hispanic and African American cohorts) significantly under-sampled in major neuroimaging datasets \cite{lim2023quantification}. 
Such lack of representation undermines the fairness and generalizability of AI models, potentially leading to biased biomarkers that overlook population-specific disease patterns. 
Consequently, this limits interpretability of neural mechanisms underlying disparities and may exacerbate inequities in diagnosis, treatment, and outcomes \cite{la2023equity, esteva2019guide}. 
To address this, it is critical to extend AI-based neuroimaging research to racially diverse populations, ensuring equitable model performance and enabling trustworthy biomarker discovery to support precision medicine across all groups.

Functional MRI–derived causal brain networks offer a principled way to quantify directional, strength-based, and valence-specific (excitatory/inhibitory) influences between brain regions.
These asymmetric interactions reflect core aspects of brain organization, and their disruption may underlie neurodegenerative and psychiatric disorders.
While interest in causal connectivity is growing, most studies focus on a few localized ROIs, limiting insights into large-scale directed network dynamics.
Such region-specific approaches overlook system-level disruptions critical to disease progression. 
Moreover, how these causal patterns vary across racial groups remains largely unexplored, leaving a significant gap in our understanding of population-specific brain–behavior mechanisms.


In this work, we propose a novel framework that models heterogeneous causal influences as directed hyper-networks among major brain systems, including seven canonical cortical networks, the cerebellum, vermis brainstem, and subcortical regions.
Grounded in the Yeo atlas and neuroanatomical priors \cite{yeo2011organization}, this coarse-grained representation enhances interpretability and reduces noise in ROI-level data.
Using these signed hyper-networks as inputs to a deep neural network, we accurately predict both Alzheimer’s disease stages and trait worry level.
Beyond prediction, we identify key directed connections that are strongly associated with different clinical phenotypes, revealing potential causal pathways involved in disease progression and mental health traits.
We also observe notable variations in the most salient directional patterns across racial cohorts, suggesting the presence of population-specific mechanisms in neurodegeneration and affective regulation.

\section{Methods}
\label{sec:methods}
\subsection{Inference of Causal Brain Networks}
To infer directed causal interactions among brain regions from fMRI BOLD signals, we employ the Independent Component Analysis–Linear Non-Gaussian Acyclic Model (ICA-LiNGAM) model \cite{shimizu2006linear, zhang2020causal}. 
Let $\mathbf{X} \in \mathbb{R}^{N \times T}$ denote the preprocessed BOLD time series, where $N$ is the number of ROIs and $T$ is the number of time points. 
ICA-LiNGAM assumes a linear non-Gaussian acyclic model:$\mathbf{x}_t = \mathbf{B} \mathbf{x}_t + \mathbf{e}_t,$
where $\mathbf{x}_t \in \mathbb{R}^{N}$ is the brain activation vector at time $t$, $\mathbf{B} \in \mathbb{R}^{N \times N}$ is the causal adjacency matrix representing directed influences among ROIs, and $\mathbf{e}_t$ is an independent non-Gaussian noise vector. 
The matrix $\mathbf{B}$ is estimated by first applying ICA to the observed signals to obtain an independent component matrix $\mathbf{S}$ and a mixing matrix $\mathbf{M}$ such that $\mathbf{X} = \mathbf{M} \mathbf{S}$. 
ICA-LiNGAM then determines a causal ordering of variables and transforms the mixing matrix into a strictly lower triangular matrix, from which the causal structure is recovered by:
$\mathbf{B} = \mathbf{I} - \mathbf{W}^{-1},$
where $\mathbf{W}$ is the estimated unmixing matrix. 
The resulting $\mathbf{B}$ captures the weighted, directed influences among brain ROIs under the assumption of linear, non-Gaussian, and acyclic data-generating processes \cite{shimizu2006linear}.
We use $\mathbf{B}$ as the directed brain network for each subject in all downstream tasks.

\subsection{System-Level Hyper-network Mapping}
To enable analysis at the macroscopic level, we construct a directed hyper-network $\mathbf{B}_{\mathrm{hyp}} \in \mathbb{R}^{K \times K}$, where each node represents one of $K$ predefined functional subsystems (e.g., Yeo’s 11 cortical and subcortical networks \cite{yeo2011organization}). 
This hyper-network is derived from the ROI-level directed causal connectivity matrix $\mathbf{B}$ and the construction involves two mappings:
\textbf{1. ROI-to-system assignment:} Each ROI $r_i$ ($i = 1, \dots, N$) is assigned to one functional subsystem $S_k$ ($k = 1, \dots, K$), based on a predefined atlas.

\noindent\textbf{2. Edge aggregation:} For each pair of systems $(S_k, S_l)$, the directed connection from $S_k$ to $S_l$ in the hyper-network is defined as the average of all directed edges from ROIs in $S_k$ to ROIs in $S_l$: $ \mathbf{B}_{\mathrm{hyp}}(k, l) = \frac{1}{|S_k||S_l|} \sum_{i \in S_k} \sum_{j \in S_l} \mathbf{B}(i, j),$
where $|S_k|$ denotes the number of ROIs in system $S_k$. Note that $\mathbf{B}_{\mathrm{hyp}}(k, l) \neq \mathbf{B}_{\mathrm{hyp}}(l, k)$ in general, preserving directionality at the system level.
This hierarchical abstraction enables interpretable system-level analysis while retaining the causal directionality embedded in the original fine-grained connectivity matrix $\mathbf{B}$.

\begin{figure}[t]
\centering
\includegraphics[width=0.48\textwidth]{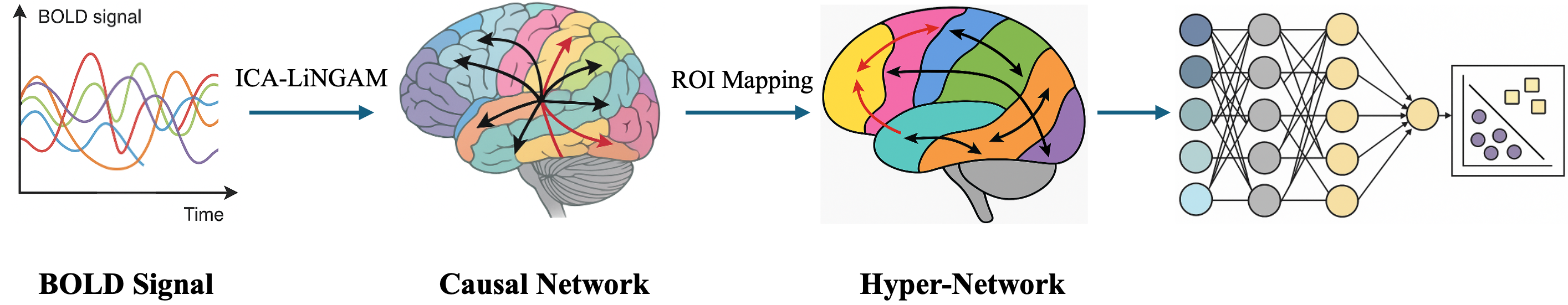}
\caption{Causal influences are inferred from ROI-level BOLD signals and aggregated into a directed hyper-network among functional subnetworks. A multilayer perceptron machine is trained to identify key hyper connectomes associated with phenotypes with SHAP.}
\label{framework}
\end{figure}

\subsection{Discovering Phenotype-Relevant Connections via Prediction Models}
To identify key system-level directed connections associated with clinical phenotypes, we flatten each subject’s hyper-network $\mathbf{B}_{\mathrm{hyp}} \in \mathbb{R}^{M \times M}$ into a feature vector $\mathbf{x} \in \mathbb{R}^{d}$, where $d = M \times (M - 1)$ is the number of directed inter-system connections (excluding self-loops). 
We train a multi-layer perceptron (MLP) model $f_{\boldsymbol{\theta}}: \mathbb{R}^d \rightarrow \mathbb{R}^C$ to predict disease stages (e.g., Normal Control (NC), Mild Cognitive Impairment (MCI), and AD) or mental health status (e.g., high vs. low trait worry). 
The model is trained using the standard cross-entropy loss: $\mathcal{L}_{\mathrm{CE}} = -\frac{1}{I} \sum_{i=1}^{I} \sum_{c=1}^{C} y_c^{(i)} \log \hat{y}_c^{(i)},$
where $I$ is the number of subjects, $C$ is the number of classes, $y_c^{(i)}$ is the one-hot encoded ground-truth label, and $\hat{y}_c^{(i)}$ is the predicted probability for class $c$.
After model training, we apply SHapley Additive exPlanations (SHAP) \cite{lundberg2017unified} to interpret the importance of each directed connectomes in hyper-network. 
SHAP assigns an additive importance value $\phi_j$ to each connectome $x_j$, such that $f_{\boldsymbol{\theta}}(\mathbf{x}) = \phi_0 + \sum_{j=1}^{d} \phi_j,$
where $\phi_0$ is the base value (expected model output), and $\phi_j$ reflects the contribution of feature $x_j$ to the prediction. 
We rank the connectomes based on their mean absolute SHAP values $|\phi_j|$ across all subjects and cross-validation folds to identify the top-K predictive directed influences. 
These connectomes offer interpretable insights into phenotype-related network changes and enable subgroup-level comparisons across racial cohorts.

\section{Results and Discussions}
\label{sec:results}
\subsection{Data Description and Implementation Details}
We used resting-state fMRI data from the publicly available Health and Aging Brain Study – Health Disparities (HABS-HD) cohort \cite{petersen2025health}.
The dataset includes 3,840 participants (mean age: $64.82 \pm 8.19$ years; 2,379 female), consisting of 1,066 African American (AA), 1,425 Hispanic, and 1,349 Non-Hispanic White adults.
Preprocessing of the functional BOLD signals was performed using the CONN toolbox \cite{whitfield2012conn}, including realignment, normalization to MNI space, spatial smoothing, nuisance regression, and band-pass filtering. 
The brain was parcellated into 132 regions of interest (ROIs) based on the Harvard–Oxford and AAL atlases and Each ROI-level BOLD series contains 150 time points.
To assess worry levels, we used the Penn State Worry Questionnaire (PSWQ), a standard measure of trait worry \cite{meyer1990development}. Consistent with prior work, individuals with PSWQ $\geq$ 60 were classified as high worry, while those below were classified as low worry \cite{molina1994penn, behar2003screening}.
Following large-scale functional system organization, our constructed hyper-network includes 11 nodes: Dorsal Attention (DAN), Default Mode (DMN), Frontoparietal (FPN), Limbic (LIM), Somatomotor (SMN), 
Ventral Attention/Salience (VAN/SAL), Visual (VIS), Cerebellum, Brainstem, Subcortical nuclei, and Cerebellar Vermis \cite{yeo2011organization, buckner2011organization}.

We performed 5-fold cross-validation in all experiments.
The model was trained using the Adam optimizer with a batch size of 32. The initial learning rate was set to 0.001 and decayed according to $((1 - \frac{\text{epoch}}{\text{max\_epoch}})^{0.9}$. We applied $L{2}$ weight decay of $1\times10^{-4}$ and a dropout rate of 0.5 to reduce overfitting. Training was early-stopped if the validation loss did not improve for 50 consecutive epochs, with a maximum of 1500 epochs. All experiments were conducted on a single NVIDIA A30 GPU.

\subsection{Race-specific Prediction of Early AD Stages}
As shown in Table~\ref{t1-disease}, classical machine learning models (SVM, Logistic Regression, Random Forest, and XGBoost) provide limited discrimination among NC, MCI, and AD, suggesting that linear or shallow decision boundaries do not sufficiently capture disease-related network structure. In contrast, the MLP trained on all subjects achieves substantially higher performance, indicating that nonlinear representations more effectively utilize functional subnetwork interactions for early-stage AD prediction.
When training the MLP separately within each racial group, performance consistently ranked highest in the White group, followed by the Hispanic and African American groups. This pattern suggests that the separability of disease stages varies across racial populations, which may reflect group-specific differences in functional network organization. Notably, the African American group is the only group performing below the pooled model, indicating that network patterns in this group may be more heterogeneous and therefore more challenging to model under a shared representation.
\begin{table}[t]
\begin{tabular}{c|ccc}
\hline
\textbf{Methods}              & \textbf{Groups}   & \textbf{Accuracy} & \textbf{Maro-F1} \\ \hline
SVM                  & \textit{All}      & 43.26$\pm$2.42             & 44.85$\pm$2.16            \\
Logistic Regres.  & \textit{All}      & 34.27$\pm$1.60             & 33.09$\pm$1.95            \\
Random Forest        & \textit{All}      & 59.24$\pm$3.39             & 58.85$\pm$2.68            \\
XGBoost              & \textit{All}      & 54.33$\pm$2.41             & 55.29$\pm$1.86            \\ \hline
\multirow{4}{*}{\textbf{MLP}} & \textit{All}      & 72.03$\pm$1.59             & 73.44$\pm$1.21            \\
                              & \textit{White}    & 85.59$\pm$1.91             & 85.01$\pm$1.77            \\
                              & \textit{AA}       & 70.52$\pm$1.89             & 70.86$\pm$0.84            \\
                              & \textit{Hispanic} & 73.10$\pm$2.04             & 74.24$\pm$1.35            \\ \hline
\end{tabular}
\caption{Mean classification accuracy and Macro-F1 scores with standard deviations under 5-fold cross-validation for NC, MCI, and AD prediction, reported for the entire cohort and for each racial group (White, African American (AA), and Hispanic).}
\label{t1-disease}
\end{table}
\begin{figure*}[htp]
\centering
\includegraphics[width=0.98\textwidth]{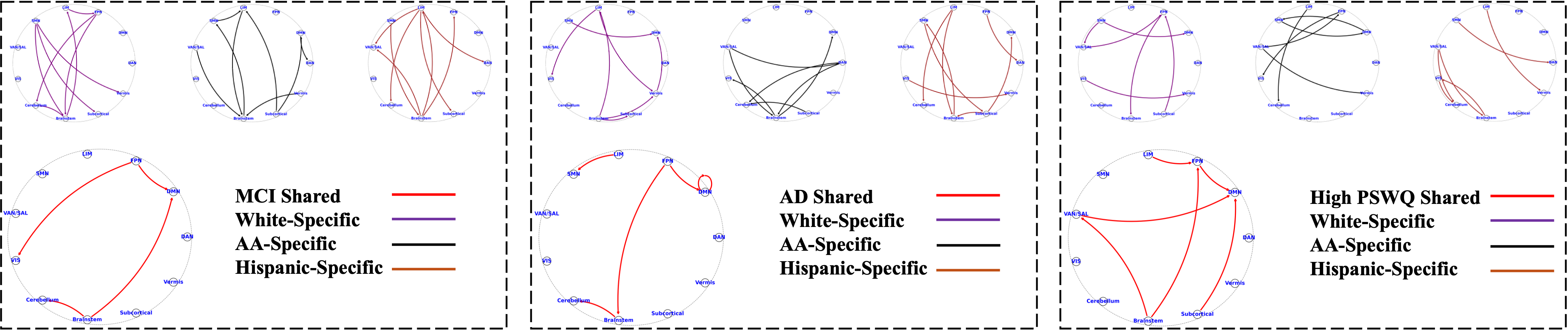}
\caption{ Top 10\% directed hyper-connectomes related to MCI, AD, and high PSWQ. The results highlight population-shared network changes (row 2) and population-specific directed connectivity patterns (row 1).}
\label{highlight-connectomes}
\end{figure*}

\subsection{Race-specific Prediction of Trait Worry Index}
As shown in Table~\ref{t2-worry}, the MLP trained on all subjects achieves substantially higher accuracy and F1 compared to other baselines, indicating that nonlinear representations are better suited for modeling worry-related differences in brain functionalities.
When models are trained separately within each racial group, all three subgroup-specific MLPs outperform the pooled MLP. 
Moreover, the performance across White, African American, and Hispanic groups remains comparably high and stable. 
This indicates that although worry-related functional patterns are present across populations, the optimal decision boundaries vary by racial group. 
Thus, race-specific modeling is necessary to fully capture the neural markers of chronic worry, rather than relying on a shared population-level representation.
\begin{table}[t]
\begin{tabular}{c|ccc}
\hline
\textbf{Methods}              & \textbf{Groups}   & \textbf{Accuracy} & \textbf{F1} \\ \hline
SVM                  & \textit{All}      & 59.65$\pm$3.72             & 61.98$\pm$2.66       \\
Logistic Regres.     & \textit{All}      & 57.77$\pm$1.59             & 56.24$\pm$1.98       \\
Random Forest        & \textit{All}      & 63.24$\pm$2.67             & 66.52$\pm$2.58       \\
XGBoost              & \textit{All}      & 69.96$\pm$2.37             & 70.96$\pm$3.16       \\ \hline
\multirow{4}{*}{\textbf{MLP}} & \textit{All}      & 84.70$\pm$1.87             & 82.79$\pm$1.42       \\
                              & \textit{White}    & 89.03$\pm$1.36             & 90.05$\pm$1.04       \\
                              & \textit{AA}       & 87.60$\pm$2.44             & 85.37$\pm$1.88       \\
                              & \textit{Hispanic} & 88.43$\pm$3.06             & 88.89$\pm$2.10       \\ \hline
\end{tabular}
\caption{Mean classification accuracy and F1 scores with standard deviations under 5-fold cross-validation for Low and High PSWQ Index prediction, reported for the entire cohort and for each racial group (White, African American (AA), and Hispanic).}
\label{t2-worry}
\vspace{-1em}
\end{table}

\subsection{Race-specific Key Hyper-connectomes}
To identify key hyper-connectome patterns that are closely associated with the target phenotypes, we performed SHAP-based attribution analysis on the trained MLP models. 
SHAP values were computed on the validation subjects of each fold and then averaged across subjects and across the 5 folds to obtain a stable importance score for each hyper-edge. 
The averaged SHAP values was subsequently ranked to identify the top 10\% phenotype-associated hyper-connectome features.
Figure \ref{highlight-connectomes} illustrates the top 10\% phenotype-associated directed hyper-connectomes identified for MCI, AD, and high-worry phenotypes. 
Statistical significance was assessed using permutation testing. 
Nearly all highlighted directed connections remained significant with $p < \frac{0.05}{11 \times 10}$, except two cases: (i) the DAN → Cerebellum connection in the African American AD group ($p = 0.0027$) and (ii) the FPN → VIS connection in the African American high-worry group ($p = 0.016$).
Figure \ref{highlight-connectomes} shows that some of these highlighted hyper-connectomes were shared across racial groups, suggesting common neurofunctional substrates relted to different phenotypes. 
For example, 4 MCI-related hyper-connectomes were consistently shared across racial groups: FPN→VIS \cite{elfmarkova2017neural}, FPN→DMN \cite{andrews2014default}, Brainstem→DMN \cite{betts2019locus}, and Brainstem→Cerebellum \cite{guell2018functional}.
In contrast to the shared hyper-connectomes observed across all groups, several connections were identified only within certain racial subpopulations. 
These findings indicate that the brain may engage different large-scale control pathways to express the same clinical or behavioral phenotype depending on context, life experience, or disease burden.
For example, within the African American high-worry subgroup, we observed a unique FPN → VIS pattern, suggesting stronger top-down monitoring of perceptual or internally imagined information, consistent with cognitive models of heightened vigilance in chronic worry. 

Figure \ref{highlight-connectomes} also indicates that in both MCI and high-worry cohorts, a consistent FPN→DMN hyper-connectome is observed across racial groups, suggesting a shared mechanism in which reduced executive control over internally directed processes contributes to both cognitive vulnerability and elevated self-focused worry.
We also find that Hispanic participants exhibit reversed directionality in brainstem–VAN/SAL interactions across the two phenotypes, brainstem→VAN/SAL in MCI and VAN/SAL→brainstem in high worry, suggesting state-dependent modulation of salience–arousal dynamics rather than a fixed regulatory pathway.
This inversion in directionality implies that the coordination between arousal systems (brainstem) and salience-driven attentional control (VAN/SAL) may shift depending on whether the dominant challenge is cognitive decline or emotional regulation. Rather than a consistent hyper-connectome signature, Hispanic participants may rely on context-specific network configurations. 

Finally, we also observed a few closed-loop directed circuits, suggesting differences in how brain networks regulate cognitive and affective states across groups.
For example the Hispanic participants with AD, a brainstem → SMN → subcortical → brainstem loop was identified, indicating that network dynamics may shift toward lower-level arousal and motor–subcortical coordination as neurodegeneration progresses.
This pattern is consistent with a reliance on more automatic and reflexive regulatory pathways when higher-order cortical support becomes compromised in later disease stages \cite{guell2018functional}.
Among African American participants with high worry, we found an SMN → DMN → SMN feedback loop, aligning with the well-established interaction between somatic tension (SMN) and self-referential rumination (DMN) in the maintenance of chronic worry \cite{paulus2006insular}.
This bidirectional coupling suggests that bodily arousal and internal thought may reinforce one another, forming a persistent emotional–somatic feedback cycle.
Importantly, these results do not imply inherent biological differences across groups, but rather highlight distinct network regulation patterns that may arise from heterogeneous life experiences, stress exposures, and sociocultural contexts.


\section{Conclusion}
This study demonstrates that directed hyper-connectomes can effectively capture functional network organization relevant to both early AD progression and worry-related behavioral traits. 
While several hyper-connectomes were shared across racial groups, reflecting common vulnerability pathways, we also identified population-specific regulatory circuits and directional interaction differences. 
These results suggest that functional network reorganization is not uniform across populations, and that incorporating population context improves predictive performance and interpretability.
Importantly, our findings highlight the value of population-aware connectomic analysis rather than a one-model-fits-all approach for neurological disease research. 

\section{Acknowledgement}
This study is partially supported by the National Institutes of Health (R21AG087888) and the National Science Foundation (CCF 2523787, IIS 2319450, IIS 2045848).
This study used data from the Health and Aging Brain Study: Health Disparities (HABS-HD).
Data collection was supported by the National Institute on Aging under Award Numbers R01AG054073, R01AG058533, R01AG070862, P41EB015922, and U19AG078109.
The content is solely the responsibility of the authors.
We also acknowledge the UTRGV High Performance Computing Resource, supported by NSF grants 2018900 and IIS-2334389, and DoD grant W911NF2110169. 

\section{Compliance with Ethical Standards}
This study analyzed de-identified human subject data from the HABS-HD study.
As the dataset is fully de-identified and publicly distributed under a data use agreement,
no additional institutional IRB approval was required.
\section{Conflicts of Interest}
The authors declare no financial or non-financial competing interests related to this work.

\small
\bibliographystyle{IEEEbib}
\bibliography{reference}

\end{document}